# Estrategia de marketing experiencial y la demanda turística en la contribución del posicionamiento de las islas flotantes Los Uros, Puno

# Experiential marketing strategy and tourism demand in the contribution of the positioning of the floating islands Los Uros, Puno

Guina Yessica Flores Montalico[1]

**RESUMEN**

El marketing experiencial, centrado en la creación de experiencias memorables y significativas para los consumidores, ha surgido como una estrategia clave en la promoción de destinos turísticos, particularmente en destinos que buscan resaltar sus características culturales únicas. En este estudio, se analiza la influencia del marketing experiencial en la demanda turística de las islas flotantes Los Uros, Puno, un destino emblemático en Perú. El objetivo de la investigación fue determinar cómo las estrategias de marketing experiencial impactan la demanda turística en dicho destino, enfocándose en las experiencias sensoriales, afectivas y de pensamiento que los turistas perciben. La metodología adoptó un enfoque cuantitativo, con un diseño no experimental de corte transversal, alcance descriptivo y un análisis correlacional. La población del estudio estuvo compuesta por 158 turistas. Para recolectar los datos, se empleó un cuestionario basado en una escala Likert de 5 puntos, con un total de 21 ítems. Entre los resultados, se encontró una correlación positiva significativa ($r = 0.809$, $p < 0.001$) entre el marketing experiencial y la demanda turística. Las experiencias sensoriales ($r = 0.780$, $p < 0.001$), afectivas ($r = 0.772$, $p < 0.001$) y de pensamiento ($r = 0.797$, $p < 0.001$) fueron descritas como factores determinantes en las decisiones de los turistas al visitar Los Uros. Además, el análisis de regresión múltiple indicó que el marketing experiencial explica el 83% de la variabilidad en la demanda turística. En conclusión, las estrategias de marketing experiencial tienen un impacto significativo en la demanda turística. Se recomienda a los encargados de marketing de la Isla Los Uros implementar tácticas que promuevan experiencias memorables, priorizando las estrategias que involucren experiencias que estimulen el pensamiento y las emociones de los turistas.

**ABSTRACT**

Experiential focused on creating memorable and meaningful experiences for consumers, has emerged as a key strategy in promoting tourist destinations. particularly in destinations seeking to highlight their unique cultural characteristics. In this study, the influence of experiential marketing on the tourism demand of the floating islands Los Uros, Puno, an emblematic destination in Peru, is analyzed. The objective of the research was to evaluate how experiential marketing strategies impact tourism demand in this destination, focusing on the sensorial, affective and thought experiences that tourists perceive. The methodology adopted a quantitative approach, with a non-experimental cross-sectional design, descriptive scope and a correlational analysis. The study population consisted of 158 tourists. To collect the data, a questionnaire based on a 5-point Likert scale was used, with a total of 21 items. Among the results, a significant positive correlation was found ($r = 0.809$, $p < 0.001$) between experiential marketing and tourism demand. Sensory ($r = 0.780$, $p < 0.001$), affective ($r = 0.772$, $p < 0.001$) and thought ($r = 0.797$, $p < 0.001$) were described as determining factors in tourists' decisions when visiting Los Uros. In addition, the multiple regression analysis indicated that experiential marketing explains 83% of the variability in tourism demand. In conclusion, experiential marketing strategies have a significant impact on tourism demand. It is recommended that marketing managers of Los Uros Island implement tactics that promote memorable experiences, prioritizing strategies that involve experiences that stimulate the thinking and emotions of tourists.

# 1 INTRODUCCIÓN

En el mundo hiperconectado de hoy, los turistas anhelan experiencias auténticas que les permitan conectarse con las culturas locales. El marketing experiencial para el turismo ofrece la oportunidad perfecta para fomentar estas conexiones, dando vida a la cultura única de un destino de una manera que genere conexiones. Este enfoque se centra en crear vivencias significativas que conecten emocionalmente con los turistas, lo que a su vez puede influir en su decisión de visitar un destino específico. Según Sakuma (2023), existe una correlación positiva significativamente entre el marketing experiencial y la fidelización de cliéntes en las medianas y pequeñas empresas en Perú, y como sugerencia indica que las estrategias de marketing que involucran la experiencia del cliente dan como resultado la lealtad de clientes y, por ende, la demanda.

La constante transformación del turismo exige brindar un enfoque más experiencial que respondan a las nuevas tendencias del mercado, en el que los turistas no solo van a visitar lugares, sino buscan vivir experiencias multidimensionales al conectar con la cultura local y sus habitantes les dejen una impresión duradera. Henche y Moreno (2018) enfatizan que el modelo turístico tradicional se transforma por alternativas que crean conexiones más significativas y memorables de tal manera satisfacen una demanda más exigente, también se interesan por experiencias integrales que contengan entretenimiento, cultura e interacción social. Esta tendencia es especialmente relevante en Perú, donde la diversidad cultural y natural ofrece un gran alcance de oportunidades para el desarrollo del sector turístico.

Las estrategias de marketing experiencial en el sector hotelero de Perú han mostrado un crecimiento del 30% en términos de inversión desde 2018, según el Informe del Ministerio de Comercio Exterior y Turismo (MINCETUR, 2023). Las estrategias de marketing experiencial en el turismo se centran en personalizar las experiencias de los huéspedes, desde la decoración inspirada en culturas locales hasta la oferta de actividades relacionadas con el entorno. El crecimiento del turismo en Perú ha impulsado la adopción de estrategias de marketing en varias regiones. Según el Informe de Turismo de PromPerú (2022), el 78% de los turistas internacionales que visitan el país buscan experiencias culturales y gastronómicas únicas, lo que ha promovido a las empresas en el rubro de turismo a diseñar un plan basado en experiencias que reflejan la autenticidad de la cultura que se promueve.

Las experiencias promoviendo la cultura vivencial en la promoción de festivales ha sido clave para el aumento del turismo interno en el Perú. En el Festival Inti Raymi, celebrado cada año en Cusco, tuvo un crecimiento en la asistencia del 12% entre 2019 y 2022, gracias a las estrategias de marketing que invitaron a los turistas a participar activamente en las celebraciones (PromPerú, 2022).

Además, luego de la pandemia COVID-19 el sector turístico tuvo la necesidad de adaptar estrategias de marketing para que los turistas puedan confiar y volver al lugar. Así mismo las empresas han tenido que desarrollar capacidades de resiliencia y adaptabilidad, implementando protocolos de higiene y seguridad para recuperar la confianza de los turistas (Hernández & Silva, 2021). Este contexto ha llevado a un aumento en la demanda de experiencias que priorizan la salud y el bienestar, lo que a su vez ha inducido a la innovación en el marketing experiencial (Ricardo et al., 2021). Por lo tanto, los lugares turísticos en Perú se vieron obligadas a considerar estas dinámicas al diseñar sus ofertas.

      El marketing experiencial también se ve involucrado con la sostenibilidad en el turismo. Por lo que se argumentan que a los diferentes destinos turísticos les convienen enfocar sus estrategias para satisfacer las nuevas exigencias de los turistas, lo cuales buscan experiencias que ademas de ser memorables deber ser también responsables y sostenibles (Mogollón et al., 2020). Esto es exclusivamente pertinente en el contexto peruano, porque se da la conservación de la biodiversidad y el patrimonio cultural los cuales son esenciales para el desarrollo turístico a largo plazo.

      Algunos estudios sobre el impacto del marketing experiencial en la demanda turística en Perú consideran las particularidades culturales y sociales del país. La riqueza cultural de Perú, donde el turista encuentra tradiciones, gastronomía y festividades, se ofrece un terreno fructífero único para el desarrollo de experiencias turísticas. El marketing experiencial puede ser una herramienta poderosa para atraer a turistas que estan en búsqueda de autenticidad (Vereda, 2023).

      Por otro lado, el marketing experiencial ha demostrado ser un enfoque efectivo para aumentar la demanda turística, especialmente en contextos como el peruano, donde la búsqueda de experiencias auténticas y memorables está en auge. Según Aguilera Sakuma (2023), existe una relación directa entre la calidad del marketing experiencial y la fidelización del cliente, lo que sugiere que un enfoque bien diseñado puede incrementar significativamente la satisfacción y lealtad del turista. Este hallazgo se alinea con estudios previos que indican que el marketing experiencial no solo mejora la satisfacción del cliente, sino que también tiene un impacto positivo en la decisión de retorno de los clientes (Ato et al., 2013).

      El turismo gastronómico en Perú ha visto un aumento del 20% en demanda entre el 2018 y 2022, según el Informe de la Cámara Nacional de Turismo (2022). Los turistas internacionales y locales buscan destinos que ofrezcan experiencias culinarias auténticas, lo que ha llevado a un

aumento en la visita a ciudades como Lima y Arequipa, conocidas por su acogida gastronómica. El impacto del turismo vivencial en la demanda de destinos rurales ha sido alto en regiones como Ayacucho y Puno. Un estudio de Paredes y Vásquez (2021) indicó que el 40% de los turistas nacionales que visitaron estas regiones en 2021 lo hicieron motivados por la posibilidad de interactuar con las personas locales y participar en las actividades tradicionales del lugar, de esa manera incrementó la demanda de los servicios turísticos en un 15%.

La inversión en infraestructura turística cuenta como un factor clave que puede influir en la demanda. Según Lahura y Sabrera (2020), las infraestructuras mejoradas llevan a un aumento significativo en el número de visitantes a Kuélap. Por lo que sugiere que, junto con el marketing experiencial, las inversiones en infraestructura son fundamentales para crear un entorno interesante que ayude a fomentar la demanda turística.

En el contexto peruano, la adaptación del marketing experiencial a las realidades locales es decisivo. Sin embargo, el estudio aborda directamente el marketing experiencial en el sector turístico, además, se enfoca en el grado de experiencias en los turistas. Según Mamani (2023) Esto es especialmente notable en un país con una amplia diversidad cultural y natural, en el que las experiencias son únicas y por ende pueden ser un diferenciador en la oferta turística.

A continuación, se muestran tres antecedentes para la investigación:

Gil-León, A., et al. (2021) en su investigación "El papel del turismo patrimonial en el índice de competitividad turística regional de Colombia: una evaluación de las relaciones mediante PLS-PM". Se evaluó cómo el turismo patrimonial contribuye a la competitividad turística en Colombia. Identificando que el turismo patrimonial posee un impacto significativo en la competitividad regional. El estudio culmina indicando que las estrategias de marketing experiencial logran potenciar el turismo patrimonial, optimizando la percepción y la demanda turística.

Seclén, J., et al. (2022) en "Marketing experiencial y el valor de marca en una empresa del sector gastronómico". Estudiaron la huella que deja el marketing experiencial en el valor de marca en el sector gastronómico. En los resultados se halló que hay un elevado valor de marca que se torna en alta demanda de compra. Este estudio enfatiza la importancia del marketing experiencial en el sector gastronómico, indicando que una experiencia positiva en el consumidor puede aumentar el valor de marca y por ende la demanda turística.

Barreto, A., & Martínez, J. (2016) en su estudio "Marketing Experiencial en FITUR: Análisis de dos destinos competidores, Islas Canarias e Islas Baleares". Estudiaron el marketing experiencial en la promoción de destinos turísticos en FITUR. En los resultados se encontró una relación positiva entre la experiencia vivida con la finalidad de visitar nuevamente el destino. En conclusión, estas experiencias positivas en ferias pueden ayudar con una influencia positiva en la decisión de los turistas, lo que es importante para el marketing en Perú.

**Variable: Marketing experiencial**

El marketing experiencial explora una comprensión profunda del marketing experiencial es una parte indispensable de la interacción con la marca actual. Ayuda a las empresas a destacarse en la feroz competencia y permite a los consumidores obtener una mayor satisfacción (Yeh et al., 2019).

El marketing es un enfoque de marketing que utiliza la experimentación e ignora las formas tradicionales de marketing. Es un enfoque de prueba y error que le permite acercarse a los clientes basándose en la experimentación en lugar del contacto directo. El principal objetivo del marketing experiencial es proporcionar un impacto muy positivo y duradero en los consumidores. (Le et al., 2018).

El marketing experiencial en el sector culinario es muy efectivo principalmente en el contexto de las PYMES y resulta muy satisfactorio en los clientes. Esto sugiere que las estrategias de marketing experiencial pueden ser altamente positivas en diversos contextos de negocio (Madiawati, 2023). El marketing experiencial busca establecer valor a través de experiencias significativas, lo que a su vez mejora la eficacia de la relación entre el consumidor y la marca. Este enfoque destaca sobre la importancia de la interacción continua y la percepción de valor en la experiencia del cliente (Lee & Peng, 2021).

El marketing destaca en el mundo de la publicidad y ofrece la oportunidad de participar en algo más significativo. Por lo tanto, cuando se ejecuta correctamente, el marketing experiencial se convierte en más que una simple transacción y se convierte en una experiencia compartida que tanto la marca como el consumidor valoran. El objetivo es dejar una impresión duradera en los participantes y en última instancia, impulsar la defensa y la lealtad a la marca (Mirando, 2023).

**Variable: Demanda Turística**

La demanda turística se refiere a la demanda de productos turísticos por parte de las personas para satisfacer su deseo de viajar con la necesidad de los individuos de desplazarse a diferentes destinos para satisfacer sus deseos de ocio, cultura y recreación (Calero, 2023)

La demanda turística está influenciada por las tendencias del comportamiento del consumidor, que buscan experiencias únicas y personalizadas, tambien, se refiere al número de turistas que desean y pueden comprar productos turísticos con una determinada capacidad de pago en moneda dentro de un período de tiempo determinado (Rendón-Contreras, 2021)

La segmentación de la demanda turística permite identificar diferentes grupos de turistas y adaptar la oferta a sus necesidades específicas (Díaz-Pompa et al., 2020). La demanda turística no solo se ve afectada por factores económicos, sino también por elementos socioculturales que moldean las percepciones y actitudes hacia el turismo (Bruna & Duque, 2019). En el análisis de la demanda turística, destacan que el pronóstico de la demanda turística es esencial para la planificación y gestión de recursos en el sector (López et al., 2021). La literatura indica que el marketing experiencial efectivo no solo mejora la satisfacción del cliente, sino que también incrementa la lealtad hacia la marca, lo que se traduce en un aumento en la demanda turística. También, se ha evidenciado que las experiencias bien diseñadas pueden influir positivamente en la percepción que los turistas tienen de un destino, aumentando así su atractivo.

***Fundamento teórico*** **entre marketing experiencial y demanda turística**

- Teoría de la Experiencia del Consumidor (Schmitt, 1999): Según esta teoría, el marketing experiencial permite que los visitantes experimenten directamente el producto o servicio brindándoles una experiencia real. La industria del turismo puede diseñar actividades de inmersión, como alojarse en una casa en un árbol o participar en experiencias culturales locales, para impresionar a los visitantes.

- Teoría de la Satisfacción del Cliente (Oliver, 1980): La satisfacción de los visitantes es clave de la demanda turística. En el contexto del marketing experiencial, cuando existe experiencias satisfactorias esto influye directamente en la percepción del turista sobre el destino, lo que agranda la probabilidad de que este recomiende el destino y a consecuencia exista mayor demanda futura.

- Teoría del Valor Percibido (Zeithaml, 1988): El valor percibido del turista se ve reflejado en la decisión de compra. En el sector del turismo, el marketing experiencial se enfoca en crear y mejorar las expectativas de los turistas para que perciban un alto valor emocional y cultural, antes de la llegada, transporte, después de la llegada y mantenimiento de la fidelidad, lo que incrementa su disposición de volver a visitar el lugar turístico y, por lo tanto, aumenta la demanda turística.

La relación entre el marketing experiencial y la demanda turística ha sido objeto de estudio, en las cuales se enfatiza que cuando las experiencias son únicas y memorables pueden repercutir en la elección del lugar de destino por parte de los turistas. Según Correa et al. (2019), el marketing experiencial se muestra como una estrategia clave para las PYMES turísticas, ya que permite vincular emocionalmente con los clientes, lo que hace que aumente la afluencia de visitantes y optimizar la rentabilidad.

Asimismo, la investigación de Mogollón et al. (2020) destaca que las empresas con mejor rendimiento dedican más tiempo al marketing experiencial, lo que demuestra la importancia del marketing experiencial. El servicio y el marketing han entrado ahora en una era en la que se valora la experiencia. Constantemente se realizan exposiciones, lo que permite al público objetivo experimentar directamente los productos y servicios, porque los consumidores obtendrán reconocimiento por los productos a través de sus propias experiencias y sentimientos reales.

En resumen, se apunta que el marketing experiencial tiene una evidencia significativa en la demanda turística, al brindar experiencias únicas que cubra la satisfacción en las expectativas que tienen los turistas, de esa manera fomentar el turismo en los destinos turísticos. El marketing experiencial para el turismo, es una estrategia de marketing dinámica que sumerge a los consumidores en experiencias de marca, ha surgido como una herramienta poderosa para atraer a los turistas potenciales y crear impresiones duraderas que prevalezcan la experiencia del turista y mantener la sostenibilidad del turismo en Perú.

Las Islas flotantes de Los Uros en Puno, a pesar de su potencial, enfrenta tasas bajas en la demanda turística, las causantes son diversas, pudiendo ser tanto la falta del desarrollo turístico sostenible o la falta de reconocimiento turístico en un mercado más amplio. Este estudio investiga el efecto de las emociones experimentadas, asi como también las percepciones cognitivas, actividades realizadas, motivación de viaje, planificación de viaje y la satisfacción, lealtad y fidelización de los turistas durante la visita a la Isla.

En un contexto donde la autenticidad y la interacción cultural profunda ganan importancia en las decisiones de los viajeros, este estudio analiza específicamente cómo se relaciona las estrategias de marketing experiencial en la demanda turística de las Islas de los Uros. Cabe mencionar que se respetara al mismo tiempo la integridad cultural y circunstancial de la Isla.

Es por esa razón que esta investigación resulta importante ya que se podrá determinar de qué manera las estrategias de marketing experiencial utilizadas en la isla influye en la demanda turística de las Islas flotantes de Los Uros. Además, con este estudio se podrá reconocer las necesidades, recordar las preferencias, entender los problemas de los turistas y de esa manera atraer y retener a los visitantes contribuyendo así al desarrollo sostenible del sector turístico las Islas flotantes de Los Uros.

## 2 MATERIALES Y MÉTODOS

### 2.1 Diseño

Para el presente estudio se utilizó un enfoque cuantitativo con un diseño no experimental, de corte transversal, alcance descriptivo y un análisis correlacional. La investigación tiene un alcance descriptivo, permitiendo proporcionar una descripción detallada y precisa de un fenómeno o población, generalmente utilizando una variedad de técnicas de recopilación de datos, través de la medición y análisis de características específicas (Danhke, 1989, citado por Hernández-Fernández & Baptista, 2004). El diseño es no experimental, porque no se manipularon las variables intencionalmente, y de corte transaccional porque la recolección de datos se realizó en un solo momento. Asimismo, el estudio es correlacional, porque estudia y analiza la relación de las variables, por tanto, busca precisar el grado de relación entre las variables las variables de marketing experiencial y la demanda turística (Hernández, 2015, p. 81).

Asimismo, Sierra (2002), menciona que un enfoque cuantitativo es el que mide la relación entre dos variables, a la vez se hace uso de la estadística con el fin de alcanzar resultados que prueben lo estudiado con herramientas, datos estadísticos, para comprobar la información que distingue entre estudios de propósito básico y aplicativo.

## 2.2  Sujetos

En este estudio, la población está conformada por 158 turistas extranjeros y nacionales que consiguen a visitar las Islas Flotantes de los Uros de la región de Puno. El muestreo es no probabilístico por conveniencia lo que indica que fue seleccionada intencional. Con una participación turística de 58.2% nacionales y 41.8% internacionales.

## 2.3  Mediciones

**Marketing experiencial y la demanda turística:** Los instrumentos utilizados en la investigación para recolectar datos fue un cuestionario basado en una escala de Likert de 5 puntos, que varía de 1 (Totalmente en desacuerdo) a 5 (Totalmente de acuerdo). El cuestionario incluyó un total de 21 ítems que abordaban las dimensiones del marketing experiencial y la demanda turística. Este instrumento es adaptado al cuestionario realizado por Pinchi (2018) en su estudio de tesis "Estrategias de marketing y su influencia en la promoción turística del parque natural de Pucallpa" en donde la confiabilidad y/o validación a través del alfa de Cronbach para evaluar la confiabilidad interna del cuestionario, fue de 0.828, lo que indica una alta consistencia interna, lo cual sugiere que los datos son adecuados para el análisis factorial y confirman la validez del instrumento.

## 2.4  Análisis estadísticos

Para el análisis estadístico se empleó el software estadístico SPSS en su versión 24.0. Los resultados descriptivos fueron presentados mediante tablas de frecuencia y porcentajes. El análisis inferencial se realizó con un nivel de significancia del 95% ($p = 0.05$) y un margen de error del 5%. Asimismo, se aplicó la prueba de normalidad de Kolmogorov-Smirnov, debido a que la población era superior a 50 trabajadores, para las variables "Marketing experiencial" y "demanda turística", junto con sus respectivas dimensiones. Como estas variables no mostraron una distribución normal ($p < 0.05$), se optó por utilizar la correlación de Rho de Spearman, una prueba no paramétrica. Por último, se aplicó un modelo de regresión lineal múltiple para analizar el efecto de la variable independiente sobre la dependiente y comprobar las hipótesis planteadas.

**2.5 Declaración sobre aspectos éticos**

La investigación ha sido avanzada con enorme responsabilidad y respetando el rigor científico y confiabilidad de información; además los turistas que participaron. fueron de manera voluntaria y quedando en anonimato para ocultar su identidad y de esa manera no comprometerlo, a su vez, las encuestas se realizaron de manera cordial, y el trato de una manera amigable. Indicando que la información será utilizada solo para fines académicos.

# 3 RESULTADOS

## 3.1 Resultados descriptivos

Según la tabla 1, se observa que la distribución por edad de los participantes es relativamente equilibrada. La mayor proporción se encuentra en el rango de 25 a 34 años, representando el 22.2%. A este grupo le siguen de cerca los participantes de 35 a 44 años y de 45 a 54 años, con un 20.3% cada uno, mientras que el grupo de 55 años a más representa el 19.6%. Por otro lado, de 18 a 24 años, constituye el 17.7%. En cuanto al género, la mayoría fueron mujeres 65.2% seguido por masculino con un 34.8%. Respecto al lugar de procedencia, se observa que el 58.2% de los participantes son nacionales, mientras que el 41.8% son extranjeros.

**Tabla 1**

*Características demográficas de la población de estudio*

| Variable | Categoría | Frecuencia | Porcentaje |
|---|---|---|---|
| Edad | 18 a 24 años | 28 | 17.7% |
| | 25 a 34 años | 35 | 22.2% |
| | 35 a 44 años | 32 | 20.3% |
| | 45 a 54 años | 32 | 20.3% |
| | 55 años a más | 31 | 19.6% |
| Sexo | Masculino | 55 | 34.8% |
| | Femenino | 103 | 65.2% |
| Lugar | Extranjero | 66 | 41.8% |
| | Nacional | 92 | 58.2% |
| | Total | 158 | 100.0% |

En la tabla 2, se presentan los resultados en términos de frecuencias y porcentajes desde la perspectiva de los encuestados sobre las variables relacionadas con el marketing experiencial y la demanda turística. Respecto a la variable marketing experiencial, la mayoría de los encuestados (67.7%) la percibe en un nivel medio, mientras que un 20.9% la identifica como alto y el menor porcentaje, un 11.4%, la considera en un nivel bajo. En cuanto a las experiencias sensoriales, el 58.9% de los participantes la ubica en un nivel medio, seguido de un 24.1% que la percibe en un nivel alto, y un 17.1% la considera baja. Respecto a las experiencias afectivas, la mayoría las sitúa en un nivel medio con un 44.3%, seguida por un 33.5% que la percibe como baja y un 22.2% como alta. En lo que respecta a las experiencias de pensamiento, el mayor porcentaje de los encuestados (47.5%) las percibe en un nivel medio, mientras que un 34.2% las considera bajas y solo un 18.4% las identifica como altas. En relación con la demanda turística, el 67.7% de los encuestados la percibe en un nivel medio, mientras que un 17.7% la sitúa en un nivel alto y un 14.6% en un nivel bajo. Por otro lado, la motivación de viaje es percibida en un nivel medio por la mayoría (63.9%), seguida de un 22.2% que la percibe como alta, y un 13.9% como baja. Finalmente, en cuanto a la satisfacción y fidelización del turista, la mayoría de los encuestados (60.1%) la percibe en un nivel medio, mientras que un 24.7% la sitúa en un nivel bajo y el menor porcentaje, un 15.2%, la considera alta.

**Tabla 2**

*Análisis descriptivo de las variables de estudio*

| Variable | Categoría | Frecuencia | Porcentaje |
|---|---|---|---|
| Marketing Experiencial | Bajo | 18 | 11,4% |
| | Medio | 107 | 67,7% |
| | Alto | 33 | 20,9% |
| Experiencias sensoriales | Baja | 27 | 17,1% |
| | Media | 93 | 58,9% |
| | Alta | 38 | 24,1% |
| Experiencias afectivas | Baja | 53 | 33,5% |
| | Media | 70 | 44,3% |
| | Alta | 35 | 22,2% |
| Experiencias de pensamiento | Baja | 54 | 34,2% |
| | Media | 75 | 47,5% |
| | Alta | 29 | 18,4% |
| Demanda turística | Baja | 23 | 14,6% |
| | Media | 107 | 67,7% |

|                                         | Alta  | 28  | 17,7%  |
|-----------------------------------------|-------|-----|--------|
|                                         | Baja  | 22  | 13,9%  |
| Motivación de viaje                     | Media | 101 | 63,9%  |
|                                         | Alta  | 35  | 22,2%  |
| Planificación de viaje                  | Media | 101 | 63,9%  |
|                                         | Alta  | 28  | 17,7%  |
|                                         | Baja  | 39  | 24,7%  |
| Satisfacción y fidelización del turista | Media | 95  | 60,1%  |
|                                         | Alta  | 24  | 15,2%  |
|                                         | Total | 158 | 100,0% |

### 3.2 Prueba de normalidad

La tabla muestra los resultados de la prueba de Kolmogórov-Smirnov, que se utilizó para evaluar la normalidad de las distribuciones de las variables relacionadas con el marketing experiencial y la demanda turística. Los resultados indican que todas las variables presentan un valor significativo (p = 0,001) inferior a 0,05, lo que sugiere que las distribuciones no siguen una distribución normal. Por lo tanto, se utilizará el coeficiente de correlación de Rho Spearman.

**Tabla 3**

*Análisis de normalidad*

| Variables | Kolmogorov-Smirnov[a] | | |
|---|---|---|---|
| | Estadístico | gl | Sig. |
| Marketing Experiencial | 0,248 | 158 | 0,001 |
| Experiencias sensoriales | 0,216 | 158 | 0,001 |
| Experiencias afectivas | 0,195 | 158 | 0,001 |
| Experiencias de pensamiento | 0,209 | 158 | 0,001 |
| Demanda turística | 0,237 | 158 | 0,001 |
| Motivación de viaje | 0,199 | 158 | 0,001 |
| Planificación de viaje | 0,206 | 158 | 0,001 |
| Satisfacción y fidelización del turista | 0,216 | 158 | 0,001 |

### 3.3 Resultados correlacionales

En la tabla 4, se observa la existencia de correlaciones significativas entre las variables del marketing experiencial y la demanda turística, todas ellas con un nivel de relación positivo y muy alto, directo y significativo (p < 0,001). Primero, el marketing experiencial presenta una correlación positiva muy alta con la demanda turística (r = 0,809, p < 0,001), de igual forma, las experiencias sensoriales presentan una correlación positiva muy alta con la demanda turística (r = 0,780, p < 0,001), asimismo, las experiencias afectivas también muestran una correlación positiva muy alta con la demanda turística (r = 0,772, p < 0,001). Finalmente, las experiencias de pensamiento presentan una correlación positiva muy alta con la demanda turística (r = 0,797, p < 0,001).

**Tabla 4**

*Análisis de correlación entre las variables de estudio*

| Variable | Demanda turística | |
| --- | --- | --- |
| | r | p |
| Marketing Experiencial | ,809** | 0,001 |
| Experiencias sensoriales | ,780** | 0,001 |
| Experiencias afectivas | ,772** | 0,001 |
| Experiencias de pensamiento | ,797** | 0,001 |

### 3.4 Resultados de Regresión

En la tabla 5, el resultado del modelo de regresión lineal múltiple, se analiza la influencia de las dimensiones del marketing experiencial sobre la demanda turística, con una variabilidad del 83%. En resumen, todas las dimensiones predictoras analizadas tienen un impacto significativo sobre la demanda turística, pero las experiencias de pensamiento son el factor más determinante, seguido por las experiencias afectivas y luego por las experiencias sensoriales. Estos hallazgos sugieren que, para aumentar la demanda turística, se deben priorizar las estrategias que involucren experiencias que estimulen el pensamiento y las emociones de los turistas.

**Tabla 5**

*Coeficiente de regresión múltiple en base a las experiencias del marketing experiencial sobre la demanda turística*

| Modelo: R2 ajustado = .83 | Coeficientes no estandarizados | | t de student | Sig |
| --- | --- | --- | --- | --- |
| | R | Error estandar | | |
| (Constante) | 12.691 | 1.712 | 7.413 | 0.001 |
| Experiencias sensoriales | 0.459 | 0.115 | 3.995 | 0.001 |
| Experiencias afectivas | 0.965 | 0.105 | 9.231 | 0.001 |
| Experiencias de pensamiento | 1.162 | 0.094 | 12.386 | 0.001 |

a. Variable dependiente: Demanda turística

# 4 DISCUSIÓN

La influencia del marketing experiencial sobre la demanda turística se ha demostrado en diversas investigaciones, destacando su excelencia en la creación de experiencias memorables que fomentan la lealtad del cliente. Una experiencia de calidad para el turista es crucial, los habitantes deben asegurarse de que cada contacto sea agradable y valioso. Cuando las marcas son capaces de ofrecer personalización e interacción instantánea, no sólo atraen nuevos clientes, sino que también hacen que los clientes existentes sean más leales a la marca. Por lo tanto, debemos prestar atención a cada detalle, porque estos pequeños esfuerzos eventualmente aportarán grandes recompensas. En un estudio realizado por Rather et al. (2021), se encontró que el marketing experiencial tiene característica con la intención de retorno de los turistas, indicando que las experiencias positivas están directamente relacionadas con la satisfacción del cliente y su pretensión de volver a visitar el destino. Yeh et al. (2019) obtuvieron resultados similares, en donde se halló que el marketing experiencial tiene una huella significativa en la satisfacción del cliente, con un $R^2$ ajustado de 0.82, lo que indica que el 82% de la variabilidad en la satisfacción del cliente. Asimismo, en el contexto del turismo cultural, Chen et al. (2022) en su estudio encuentran que la percepción del marketing experiencial tiene un impacto efectivo en las emociones del consumidor, con un $R^2$ ajustado de 0.79. Este estudio enfatiza cómo las conexiones emocionales que se generan a través de experiencias bien diseñadas pueden influenciar en la percepción del destino y en la decisión de los turistas a la hora de elegir un destino turístico. La investigación propone que los destinos que establecen estrategias de marketing experiencial centradas en la cultura pueden mejorar su competitividad y atraer una mayor demanda turística. Sousa & Alves (2019) dejan ver que el marketing experiencial y la calidad de la experiencia son definitivos en las intenciones de comportamiento de los turistas, con un $R^2$ ajustado de 0.85. Este hallazgo destaca la necesidad de que los destinos turísticos, especialmente en nichos como el turismo, se enfoquen en crear experiencias memorables donde no solo se cumpla las expectativas de los turistas, sino que también fomenten relaciones duraderas. Finalmente, el estudio de Lee & Peng (2021) encontró que el marketing experiencial tiene efectos positivos directos en el valor experiencial y la lealtad del cliente, con un $R^2$ ajustado de 0.78. Este estudio termina la literatura existente al proporcionar evidencia empírica de que las actividades de marketing experiencial pueden acrecentar las emociones de los consumidores, lo que a su vez perfecciona su intención de regresar. Estos resultados refuerzan la idea de que el marketing experiencial es una estrategia efectiva para incrementar la demanda turística, al crear vínculos emocionales que motivan a los turistas a volver.

La percepción del marketing experiencial en el contexto turístico, que se ha señalado en un 20.9%, manifiesta una tendencia gradual hacia la búsqueda de experiencias memorables por parte de los turistas. Aguilera Sakuma (2023), en su estudio sobre marketing experiencial y lealtad del cliente entre sus resultados encuentra que un 22% de los encuestados valoran de manera positiva las experiencias ofrecidas por los habitantes de Los Uros. Este resultado resalta la necesidad de que se implementen estrategias que prevalezcan la creación de experiencias memorables para atraer mayor demanda turística. Asimismo, el análisis de la percepción de los servicios turísticos en Cuenca, Ecuador, realizado por Mayorga Ases (2023), concluye que un 19.5% de los turistas valoran positivamente la administración de los servicios turísticos en la ciudad. Este resultado es semejante al 20.9% donde sugiere que, a pesar de las diferencias contextuales, la percepción del marketing experiencial es aún un aspecto decisivo para la satisfacción del visitante. Finalmente, el estudio de Juarez Reto (2023) sobre la relación entre el marketing experiencial y la calidad del servicio en una empresa de repuestos eléctricos , los clientes valoraron positivamente la experiencia de compra con un 20%, lo que fortifica la idea de que la percepción del marketing experiencial es un factor clave en la satisfacción del cliente, lo que involucra que deben enfocarse en crear experiencias no solo memorables, sino que reconozcan las expectativas y necesidades de los turistas.

Finalmente, la relación entre el marketing experiencial y la demanda turística sobresale la importancia de las experiencias en la decisión de los turistas al elegir un destino. Resultado que propone que las estrategias que priorizan la creación de experiencias memorables pueden aumentar la satisfacción del cliente y, por ende, influir en su decisión de visitar un destino turístico. Además, el estudio apoya a la idea de que el marketing experiencial tiene un impacto significativo en la percepción del cliente y, por ende, en la demanda turística. Seclén et al (2022)

En su investigación, se reportó que un 70% de los consumidores valoran positivamente las experiencias ofrecidas por las marcas, lo que se traduce en un aumento en la intención de compra y lealtad hacia la marca. Cueva Vega (2023) En su estudio, encuentra que el 90% de los encuestados consideran que las experiencias únicas ofrecidas en la región son un factor determinante para su decisión de visitar un lugar turístico. En conclusión, con la obtención de los resultados de este estudio se sugiere que las estrategias de marketing experiencial pueden ser un motor clave para el desarrollo económico local, al atraer a más turistas y generar empleo en la comunidad. Se refuerza la idea de que el marketing experiencial tiene un efecto positivo en la competitividad en el sector turístico.

## 5    CONCLUSIÓN

La conclusión sobre el objetivo principal se evidenció que existe una influencia significativa entre las estrategias de marketing experiencial y la demanda turística, con un $R^2$ ajustado de 0.83 y $p < 0.001$. Estos resultados son consistentes con estudios recientes que también han encontrado correlaciones positivas entre el marketing experiencial y la satisfacción del cliente en el sector turístico. Sugiriendo que el marketing experiencial es fundamental para mejorar la satisfacción y, por ende, la demanda turística. Este hallazgo fortifica la idea de que las experiencias memorables no solo cautivan a los turistas, sino que también fomentan su lealtad y deseo de regresar. Chen et al. (2022) complementa estos resultados al probar que el marketing experiencial tiene un impacto positivo en la emoción del consumidor, lo que a su vez tiene influencia en su intención de regresar a destinos culturales. La conexión emocional que se crea a través de experiencias memorables es un factor clave que puede establecer la percepción del destino y la decisión de los turistas de volver, lo que se forma con los resultados de esta investigación. Finalmente, destacamos que los resultados esta investigación, adyacente a la literatura existente, resaltan la importancia del marketing experiencial como una herramienta efectiva para cautivar y retener turistas, lo que es fundamental para el desarrollo sostenible del sector turístico en Perú y a nivel global.

Respecto al objetivo dos, la Isla Los Uros, es un destino turístico emblemático que ha experimentado un crecimiento significativo en la demanda turística gracias a su singularidad cultural y natural. Sin embargo, la percepción de marketing experiencial en este destino se ha establecido en un 20.9%, lo que indica que, aunque hay un reconocimiento de la importancia de las experiencias ofrecidas, aún existe un margen considerable para mejorar la efectividad de las estrategias de marketing. Estos resultados indican que existe una carencia que implementando estrategias podría beneficiarse de un enfoque más robusto en la creación de experiencias memorables que conecten emocionalmente con los visitantes. No obstante, de que un 20.9% de los encuestados evalúa positivamente el marketing experiencial, la demanda turística del 17.7% muestra que no todos los turistas son alcanzados por estas estrategias. Finalmente, los resultados de esta investigación recalcan la importancia de un enfoque centrado en el cliente en el marketing experiencial. La implementación de experiencias en Los Uros no solo debe ser memorables, sino que también deben responder a las necesidades y deseos de los turistas, es fundamental para acrecentar la demanda. Al incorporar el marketing experiencial en las estrategias, el encargado de marketing no solo optimizara la percepción de sus expectativas, sino que también provocara una mejor lealtad y satisfacción entre los visitantes, lo que es trascendental para el crecimiento sostenible de este sector turístico en el futuro.

En conclusión, la relación significativa entre el marketing experiencial y la demanda turística, con un coeficiente de correlación de $r = 0.80$, $p < 0.005$. Confirma que las estrategias apoyadas en experiencias juegan un papel primordial en la persuasión y fidelización de turistas, incrementando así la demanda de destinos turísticos. La relación encontrada destaca la importancia de seguir transformando en experiencias memorables que contribuyan positivamente en las decisiones de los turistas. La consistencia en los resultados propone que las experiencias trazadas estratégicamente pueden impresionar positivamente en la competitividad de los destinos. Finalmente, este estudio recalca la necesidad de que los encargados de marketing de Los Uros prevalezcan el marketing experiencial como parte central de sus estrategias de crecimiento. Al ofrecer experiencias únicas y personalizadas, es posible no solo atraer a más turistas, sino también generar una demanda sostenida en el tiempo, lo que contribuye a la sostenibilidad y éxito de los destinos turísticos.

# 6   REFERENCIAS